\def\year{2018}\relax
\documentclass[letterpaper]{article} 
\usepackage{aaai18}  
\usepackage{times}  
\usepackage{helvet}  
\usepackage{courier}  
\usepackage{url}  
\usepackage{graphicx}  
\usepackage{color}
\usepackage{balance}
\begin{document}
%
\title{An Experimental Study of Structural Diversity in Social Networks}
\author{
Jessica Su\\Facebook, Inc.\thanks{Work done when the author was at Stanford University.}
\And
Krishna Kamath\\Twitter, Inc.
\And
Aneesh Sharma\\Google, Inc.\thanks{Work done when the author was at Twitter, Inc.}
\And
Johan Ugander\\Stanford University
\And
Sharad Goel\\Stanford University
}
\maketitle
\begin{abstract}
Several recent studies of online social networking platforms have
found that adoption rates and engagement levels are positively correlated 
with \emph{structural diversity}, the degree of heterogeneity among an individual's 
contacts as measured by network ties. One common theory for this observation is that 
structural diversity increases utility, in part because there is value to interacting 
with people from different network components on the same platform.
While compelling, evidence for this causal theory comes from observational studies, 
making it difficult to rule out non-causal explanations.
We investigate the role of structural diversity on retention
by conducting a large-scale randomized controlled study on the Twitter platform. 
We first show that structural diversity correlates with user retention on Twitter, 
corroborating results from past observational studies.
We then exogenously vary structural diversity by altering the set of network
recommendations new users see when joining the platform; we confirm that this 
design induces the desired changes to network topology.
We find, however, that low, medium, and high structural diversity treatment 
groups in our experiment have comparable retention rates.
Thus, at least in this case, the observed correlation between structural diversity 
and retention does not appear to result from a causal relationship,
challenging theories based on past observational studies.
\end{abstract}

From farmers adopting agricultural methods~\cite{ryan1943diffusion} to runners deciding to exercise~\cite{aral2017exercise}, 
people often consider the actions of their 
social contacts when making decisions~\cite{coleman1957diffusion,bass1969new,schelling1971dynamic,granovetter1978threshold,bikhchandani1992theory,watts2007influentials,bakshy2012social}.
A basic theme in this literature
is that an individual's likelihood to take an action
increases with the number of their contacts who have taken that action.
For example, those with more friends who exercise are more likely to exercise themselves~\cite{aral2017exercise}. 

The recently proposed theory of structural diversity~\cite{ugander2012structural} enriches this core idea by suggesting that individuals consider not merely the \emph{number} of contacts who have engaged in some action, but also the \emph{network structure} of those individuals (i.e., the social relationships between them). 
There are multiple ways to quantify structural diversity, but the original and perhaps simplest measure is the number of connected components among one's contacts who have adopted.
According to this theory, an individual is more likely to exercise if two college friends and two work colleagues exercise than if four college friends exercise but no work colleagues do.
Several observational studies are consistent with the structural diversity hypothesis, finding that one's propensity to act is predicted by the structural diversity of one's contacts who have acted~\cite{spiliotopoulos2013understanding,weng2013virality,backstrom2013characterizing,aral2017exercise}.

There are compelling theoretical arguments for structural diversity affecting behavior.
For example, 
individuals might prefer the diversity of opinions expressed on an online platform that includes people across their social circles over the narrowly tailored
content on a service that caters to a single group~\cite{page2008difference}.
Further, individuals may personally benefit  from mediating relationships between otherwise disconnected groups \cite{burt2000network}.
Finally, with high structural diversity one might be more likely to encounter the single individual or group whose actions are influential, with diversity thus acting as a hedging mechanism:
if some individuals would join a service only if their family members have joined, and others only if their friends have, we would expect to see adoption increase with structural diversity.

These explanations all suggest that exogenously increasing structural diversity would
increase adoption, retention, and engagement.
There is, however, an alternative class of purely correlational explanations for why adoption, retention, and engagement would be associated with increased structural diversity.
For example, if retention is exclusively a function of age, and younger individuals have more structurally diverse networks,
structural diversity would correlate with retention despite being incidentally related to it.
A more subtle example stems from the popular ``strength of weak ties'' theory~\cite{granovetter1973strength,aral2014tie} of information flow. 
Suppose individuals with structurally diverse contact networks are socially farther from the population of existing adopters than those with structurally homogeneous networks.
Then, if non-adopters in the core of the network (with homogeneous networks) have actively chosen to abstain, while those on the periphery (with diverse networks) are still deciding whether to adopt, we would expect structural diversity to correlate with adoption---but there would again be no causal connection.

Past studies of structural diversity have all been observational or quasi-experimental, and so it has been difficult to disentangle causation from correlation.
As we describe below, experimental tests of structural diversity must address
serious theoretical challenges in defining the causal estimand of interest, as well as
difficult logistical hurdles in carrying out a suitable experiment at scale---two issues that have received relatively little attention in the literature.

After discussing these high-level concerns, we detail our general approach, and then describe the results of a randomized controlled experiment on the Twitter platform that directly tests for a causal link between structural diversity and retention.
In line with results from past observational studies, we first show that structural diversity indeed correlates with user retention on Twitter.
We then exogenously vary structural diversity by altering the set of network recommendations new users see when joining the platform.
We show that this randomization strategy induces  the  desired  changes  to  network  topology.  
However---and this is our key result---low, medium, and high structural diversity treatment groups in our experiment have comparable retention rates. 
Thus, at least in this case, the correlation we observe between  structural  diversity  and  retention  does  not  appear  to result from a causal relationship.
Our results accordingly challenge causal explanations of 
the link between structural diversity and behavior
that stems from past observational studies, and highlight 
the importance of randomized experiments in network analysis. 

\section{Related Work}
The role of network structure in social decision making is at the foundation of the vast literature on the diffusion of innovations, and many studies have examined this relationship. \cite{coleman1957diffusion,rogers1962diffusion}.
The growth of the internet and online services over the last 20 years have provided a particularly well-instrumented and fruitful research environment for studying decision making on networks. 
The present work is motivated by the fact that most prior studies leveraging the opportunities of online data are observational, and thus can generally only provide correlational---not causal---evidence for the role of network structure in decision making.

Our focus in this work is on the role of structural diversity in new user engagement, in line with the original paper on structural diversity \cite{ugander2012structural}. 
Beyond user engagement, related work on prediction problems involving network diversity have found evidence that more diverse networks among Twitter hashtag participants~\cite{weng2013virality,romero2013interplay}, Sina Weibo message distribution networks~\cite{bao2013popularity}, and Facebook conversation thread participants~\cite{backstrom2013characterizing} all predict greater growth of the hashtag usage, message distribution, and conversation, respectively.

A recent study of decisions to exercise through a running application provides important causal evidence of social influence between friends~\cite{aral2017exercise}. That study examines the role of structural diversity as part of its work, and finds that structural diversity is correlated with more social influence. While the estimates of social influence are causal---achieved through an instrumental variables approach using weather as an instrument---the evidence for a relationship between the outcome and structural diversity is only correlational.

Further, recent work examining cellular phone networks in a south Asian country found that structural diversity, measured as the fraction of open triads in an ego-network, is positively correlated with individual income ~\cite{jahani2017differential}. This finding highlights the importance of understanding demographic correlates of structural diversity that may explain findings that previously have been understood as being caused by structural diversity.

Finally, we highlight an early examination of the role of network structure in the adoption of online products that studied users choosing to join groups within the blogging service LiveJournal \cite{backstrom2006group}. This study found that a high edge density within a user's friends that were already in a group---which could be interpreted as a contact neighborhood with low structural diversity---predicted that they would join that group. This predictive finding is consistent with the idea of users valuing social cohesion \cite{burt1987social} in choosing groups;
but it is in opposition to the role of structural diversity identified in the work discussed above. 
A subsequent  meta-analysis of online social applications reported a mixture of positive and negative predictive relationships between edge density and adoption~\cite{kloumann2015lifecycles}.

\section{Two Challenges for Experimental Studies\\of Structural Diversity}

Any experimental study of structural diversity faces two key challenges, one conceptual and the other operational.
First, it is conceptually unclear how even to define the causal effect of interest in this case.
Typically one investigates the causal effect of a specific treatment, like taking a particular drug, and
to make progress one must make assumptions that exclude certain causal pathways, like assuming the temperature outside does not affect a patient's response.
These assumptions may not hold exactly, but they are often relatively weak and acceptably approximate reality. 
However, experimental tests of complex social phenomena typically require comparatively strong assumptions~\cite{sobel:2006}.
We could, for example, assume that adoption is affected only by the degree of structural diversity in an individual's network---and exclude the effect of all other network features.
But this assumption is unrealistic:
a contact network comprised of work colleagues and college friends is likely very different than one comprised of family members and one's bowling league, even if the two have comparable structural diversity.
In general, if one believes that network membership---and not just network structure---matters, standard exclusion restrictions are hard to justify.

In light of this limitation, one might consider holding constant the members in one's contact network, and then study what happens as connections between those members vary. 
While theoretically interesting, this formulation encapsulates a somewhat narrow 
view of structural diversity---and also one that is particularly difficult to test.\footnote{%
Arguably, changing an individual's network connections also changes that individual's persona, so even this formulation cannot completely escape the entanglement of structure and membership.
}
The common causal explanations for the structural diversity phenomenon
are mediated by membership, and hence do not admit a natural separation of membership and structure.

The second, operational challenge is to 
exogenously induce structural changes to an individual's network of social contacts.
Random team assignment has been used to alter network membership~\cite{hasan:2017}, and random roommate assignment to alter single dyads in student dormitory populations~\cite{lyle2007estimating}.
But studying structural diversity inherently involves much larger changes to network structure, and studying the consequences of such changes requires running experiments at scales that are typically infeasible with traditional designs.

\section{Experiment Design}

One cannot completely circumvent the conceptual and operational obstacles identified above. In this work, however, we were able to mitigate them by carrying out the following randomized experiment on the Twitter social network.

Our study population consists of a sample of 4.2 million new users who joined Twitter during a particular period in 2016. 
Users were selected into the sample at random from among all new users who joined during the study period, and so our study population is approximately representative of new Twitter members during 2016.

Our outcome variable of interest is 3-month retention, 
meaning that the user visited Twitter at least once in the 30-day period 90--120 days after joining. 
Our primary goal is to understand the causal relationship between structural diversity and retention defined in this manner.
This measure of activity is a common benchmark---both at Twitter and elsewhere---though we note that our results are qualitatively similar if we look at related metrics (e.g., 6-month or 12-month retention).

\subsection{Defining structural diversity}
In designing our experiment and evaluating the results,
we use a variant of average pairwise cosine similarity~\cite{sharma:2017}---defined below---as our measure of structural diversity.
With this measure, 
lower similarity corresponds to higher diversity. 
Though our definition of structural diversity differs somewhat from the original component-based metric~\cite{ugander2012structural}, it facilitates finer-grained analysis, particularly in larger contact networks where component counts can be overly coarse.

To start, we define the similarity sim$(v,w)$
for any two users $v$ and $w$.
Specifically, sim$(v,w)$ is the cosine similarity of the binary incidence vectors for $v$ and $w$, where these vectors indicate which users are following them.
That is,
\begin{equation}
\textrm{sim}(v,w) = \frac{A \cdot B}{||A||\cdot||B||},
\end{equation}
where $A$ and $B$ are the incident vectors for $v$ and $w$, respectively.
Thus, $v$ and $w$ are considered similar if they are followed by a similar set of users.

In theory, to find the average pairwise similarity for a set of users, we would compute and average the similarity scores over all pairs of users in the set. 
However, due to limitations in Twitter's production system, for any user $v$ we are only able to compute similarity scores for the 40 users most similar to $v$. We thus set sim$(v,w) = 0$ for $w$ outside this set of 40 when computing the average.\footnote{In order to use similarity scores that reflect the current state of the graph at the time that the user is signing up, we use the near-real time similarity scores available through a caching service with a limited capacity.}

\begin{figure}[t]
\centering
\includegraphics[width=0.85\columnwidth]{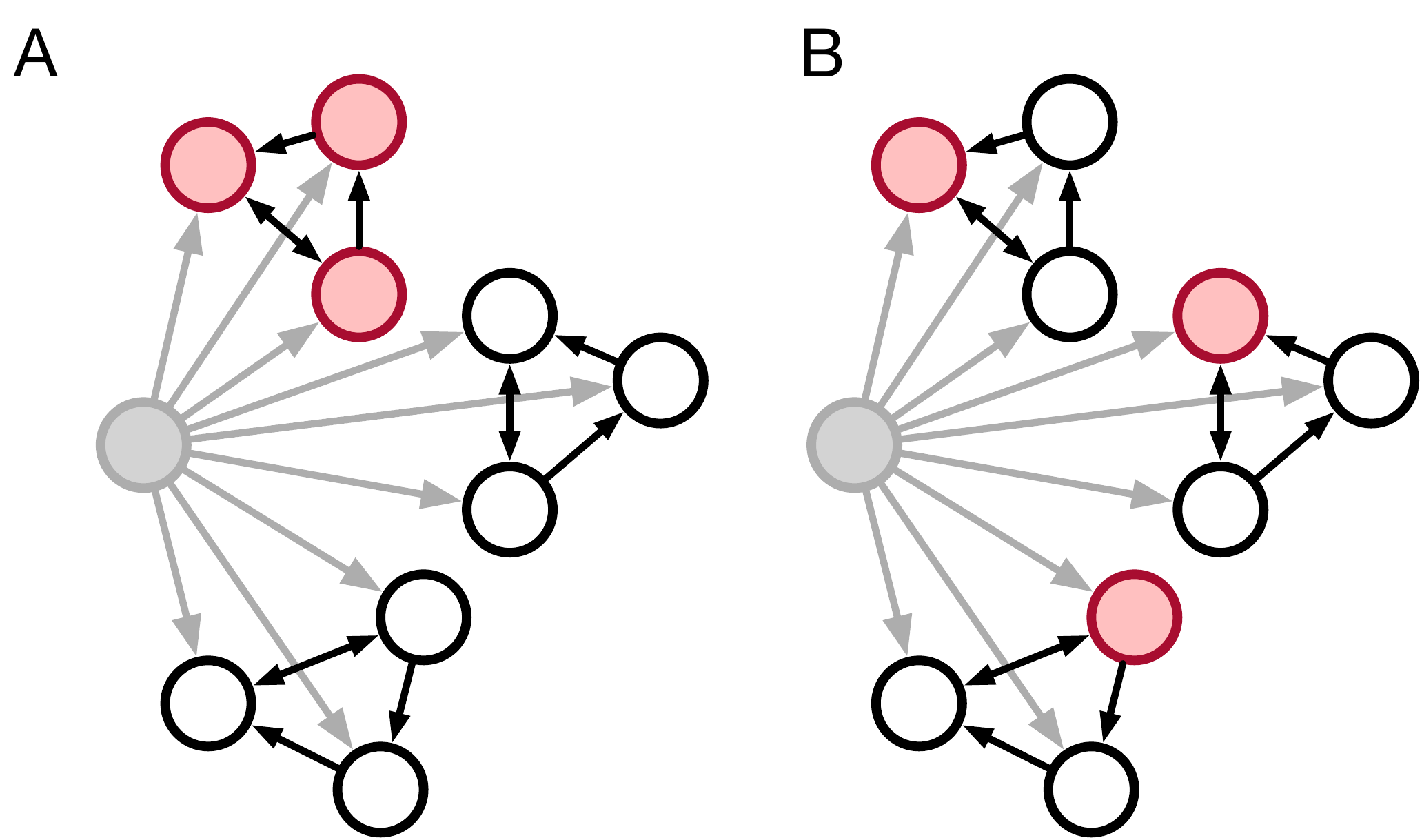}
\caption{In a stylized example, a new user (gray node) has 9 candidate recommendations connected by a similarity graph consisting of three weakly connected components. The goal is to recommend 3 users.
(A) To choose a high-similarity (low-diversity) set, we select three nodes (red) from the same component.
(B) To choose a low-similarity (high-diversity) set, we select one node (red) from each weakly connected component in the similarity graph.}
\label{fig:components}
\end{figure}

\subsection{Randomizing structural diversity}
We now describe how we exogenously varied structural diversity to test for a causal link between diversity and retention.
Users who create an account on Twitter are taken through an onboarding process, and account recommendations for the user are a key part of this process.
The onboarding process evolves continually, but at the time of running our experiment, new users were shown recommendations on two different pages. First, users were asked to select topics
that they might be interested in (e.g., ``Science \& Technology''), and were then shown recommendations related to those topics (e.g., the accounts of well-known scientists).
Users would bypass this screen if they did not select any topics of interest. Then, the user was given the choice to import their phone and email contacts. Whether a user chose to import their contacts or not, they were taken to a second screen where they were given another set of account recommendations, this time focused on their social circle. Our experiment only affected the recommendations on this second screen, which we term the set of ``contact recommendations''.

The contact recommendations shown to a user were based on a machine-learned ranking that took a base set of candidates, and ranked them by probability of generating friend (bi-directional follow) relationships. The set of candidates was generated by taking the list of phone and email contacts for the user, as well as a subset of their topical follows from the first screen\footnote{Especially if they had no existing contacts on Twitter.}, and running a SALSA-like algorithm on the friend graph of these users \cite{gupta:2013,su:2016}. 
In essence, contact recommendations were a mix of a user's direct phone/email contacts and friends of these contacts. Since these are arguably social contacts, our diversity experiment focused on these in order to stay close to the prior work on diversity, which is mostly on social networks.

\begin{figure*}[t]
\centering
\includegraphics[height=7cm]{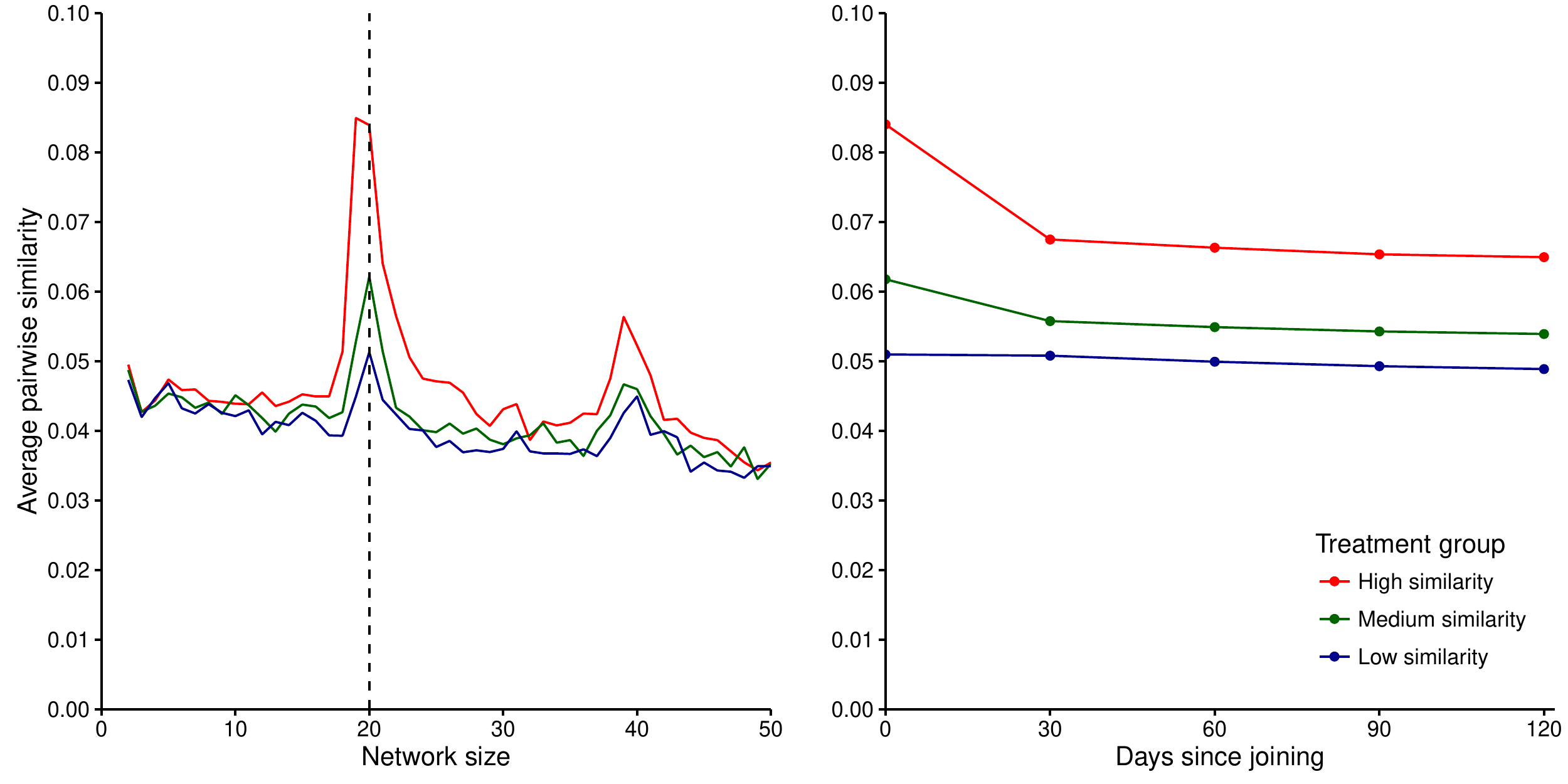}
\caption{Left: Among those users who followed all our 20 recommendations (i.e., the ``follow all'' users), the treatments induce significant changes to structural diversity, as measured by average pairwise similarity
computed on a user's initial social network.
Right: Average pairwise similarity over time, by treatment condition, for the ``follow all'' users,
where similarity is based on the network at the point in time indicated on the horizontal axis.}
\label{fig:survivors}
\end{figure*}

\begin{figure*}[t]
\centering
  \includegraphics[height=7cm]{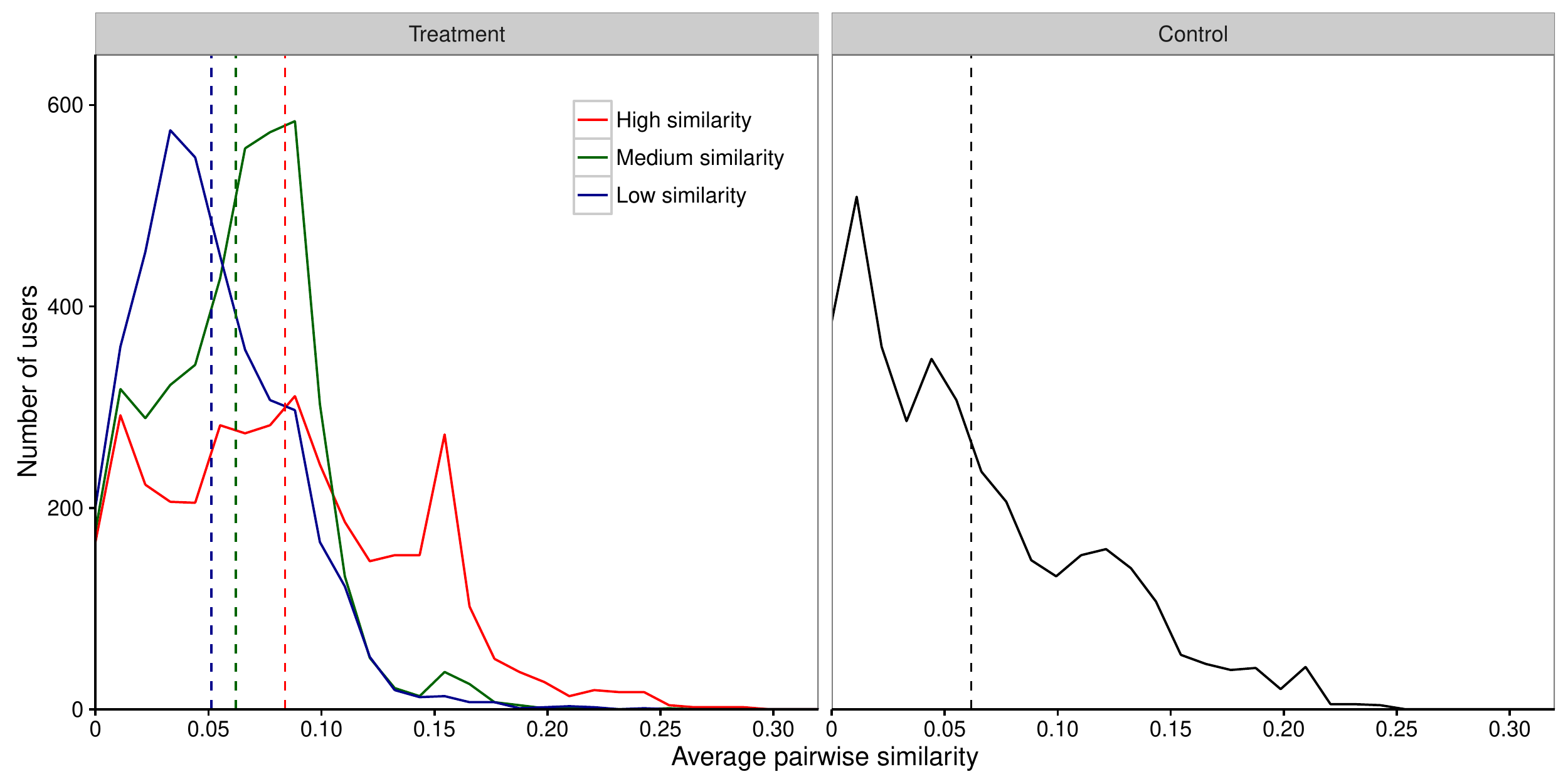}
\caption{The distribution of average pairwise similarity (based on a user's initial social network) across users in each treatment group, for users who followed exactly 20 people.  The dashed lines mark the mean for each treatment group.}
\label{fig:sim-dist}
\end{figure*}

\begin{figure*}[t]
\centering
 \includegraphics[height=7cm]{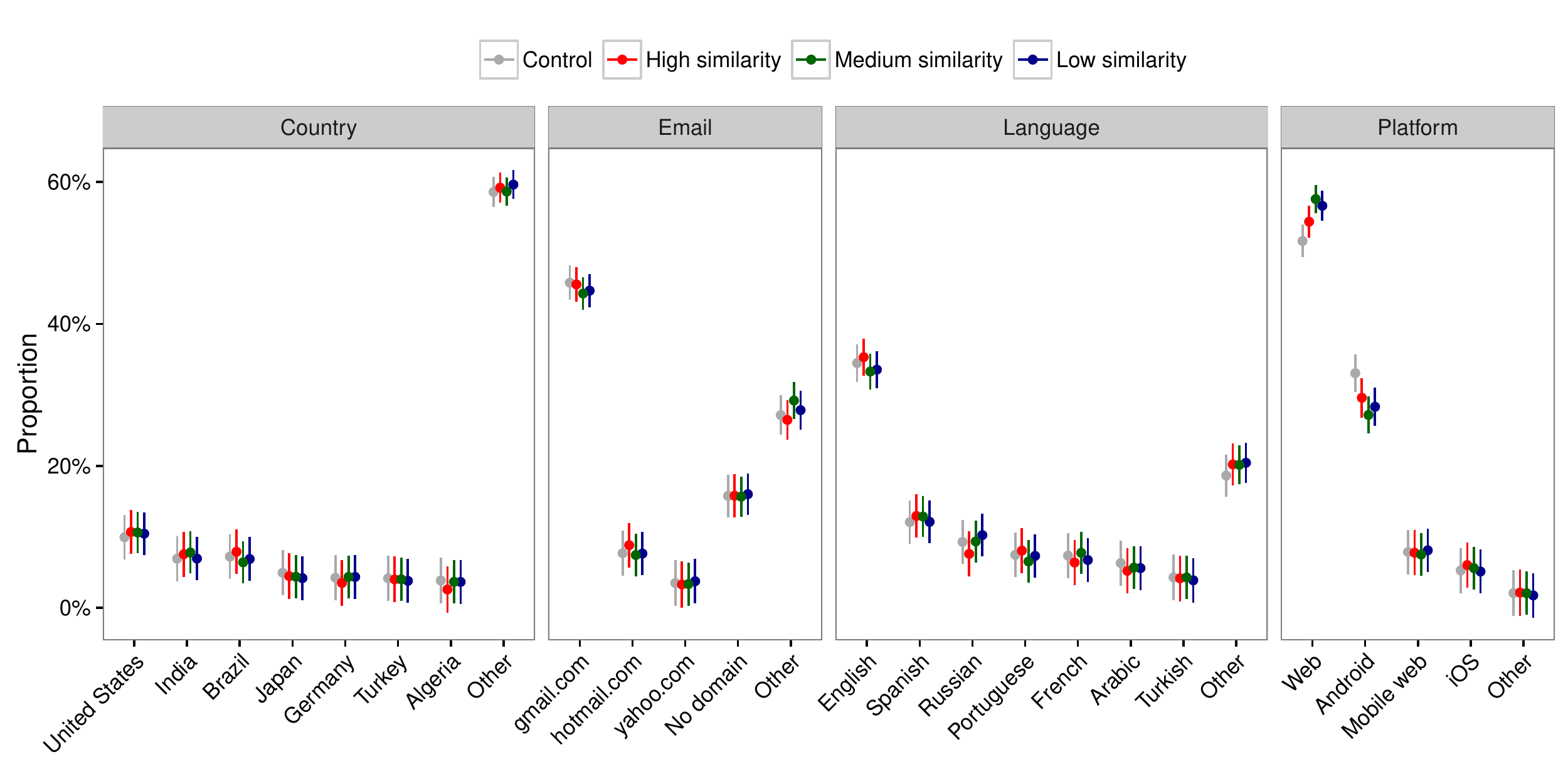}
\caption{Covariate distributions across treatment conditions for the subpopulation of users who followed all our 20 recommendations.
Treatment arms have nearly identical distributions, mitigating concerns of selection bias due to conditioning on the post-treatment ``follow all'' behavior.
}
\label{fig:balance}
\end{figure*}

From among the dozens of personalized contact recommendations $C$ generated for each new user by Twitter's internal algorithm, we attempted to select three different subsets of 20 users to promote: subsets with low, medium, and high structural diversity, as operationalized by average pairwise similarity.
Among these candidates, we first considered the induced ``similarity graph'' $G_C$ on the candidates, where we drew a directed edge from user $u$ to user $v$ 
if $v$ was one of the 40 most similar users (measured by cosine similarity, as discussed above) to $u$ at the time of the experiment.

Now, given the directed similarity graph $G_C$, we then found the three largest weakly connected components in $G_C$.  In order to be deemed \emph{eligible} for the experiment, 
we required that a user have at least one component of size $\geq 20$, two components of size $\geq 10$, {\it and} three components of size $\geq 7$.
(This restriction was enforced so that 
we could induce the desired structural changes, as described below.)
If a user was deemed eligible for the experiment, we assigned the user uniformly at random to one of four treatment conditions: 
``control,'' ``high similarity,'' ``medium similarity,'' or ``low similarity''.

In the ``high similarity'' condition, we recommended the 20 most relevant users from the largest component, where relevance was based on Twitter's internal ranking. 
In the ``medium similarity'' condition, we recommended the 10 most relevant users from the largest component and the 10 most relevant users from the second-largest component. 
In the ``low similarity'' condition, we recommended the 7 most relevant users from each of the two largest components, and the 6 most relevant users from the third-largest component. 
This design is based on the idea that recommendations chosen from different connected components are likely to be more diverse
than those chosen from the same connected component.
Figure~\ref{fig:components} shows a
stylized visualization of this strategy.

In the ``control'' condition, 
we displayed the users that Twitter would have ordinarily shown (i.e., we did not alter the standard recommendations).
We note that users in the ``control'' group were eligible users,
and thus could have been randomly assigned to any of the other three conditions.
As a result, we can compare ``control'' users to ineligible users to understand differences between eligible and ineligible populations that all receive the standard, unaltered recommendations.

Approximately 500,000 users (12\% of our study population of 4.2 million)
were ultimately deemed
eligible for the experiment.
That is, 
we could construct low-, medium-, and high-similarity recommendation sets 
for these 500,000 users.
As an example of when this set construction failed,
we generally could not create low-similarity (i.e., high-diversity) sets from the recommendations generated for users with few known social contacts on Twitter, as most recommended users in this case were in a single, large connected component.

After constructing these recommendation sets, we randomly assigned each eligible user to one of four treatment arms (low, medium, high, or control), and then promoted
that arm's subset of 20 users to the top of the recommendation list; in the control condition we did not alter the standard recommendations.
Recommendations were shown to eligible users 20 at a time, and they were free to follow as many or as few of the recommendations as they wished.

As our experiment involves human subjects, we sought (and received) approval from Stanford University's Institutional Review Board. 
We note that our intervention consisted only of altering the set of recommendations presented to new Twitter users---with these recommendations chosen from Twitter's standard personalized set of candidates. Our experiment protocol was deemed to pose minimal risk to users, and was thus evaluated under expedited review.

\section{Assessing the Assignment Mechanism}

\subsection{Induced structural changes}

Our experimental design is not guaranteed to induce the desired changes to network structure; 
even if we encourage new users to follow a structurally diverse set of people, they might nevertheless elect to follow a low-diversity set.
Some users, however, opted to follow exactly the 20 users we selected, in part because of a ``follow all'' option that appeared on the recommendation page.
Among these ``follow all'' users, Figure~\ref{fig:survivors} (left) shows that our treatment conditions indeed induced significant changes to structural diversity. (The peak at 40 is due to a similar effect, with users accepting all suggestions on another page of recommendations.
The number of users following 40 recommendations is small---less than 1\% of eligible users---and we do not consider them as part of our population of ``follow all'' users.)
In contrast, among those who ultimately chose to follow 30 users, the structural diversity of their resulting networks did not vary substantially across our three treatment conditions, because they followed a combination of accounts we recommended and those that we did not, hampering our ability to measure the causal effect of structural diversity on retention.

Among follow-all users, the high-similarity (low-diversity) group has mean pairwise similarity that is 64\% higher than the low-similarity (high-diversity) group (0.084 vs.\ 0.051),
and Figure~\ref{fig:survivors} (right) shows that the induced gap in structural diversity persists over time.\footnote{%
For follow-all users, 
the 95\% confidence interval on mean similarity for the high-similarity group is 
$[0.075, 0.093]$;
for the medium-similarity group it is 
$[0.055, 0.070]$; and
for the low-similarity group it is 
$[0.044, 0.058]$.
In particular, the three treatment groups are generally well separated.
}
Figure~\ref{fig:sim-dist} shows the full distribution of pairwise similarity for follow-all users in the low-, medium-, and high-diversity treatment arms, 
indicating that there is significant heterogeneity within each treatment arm, as one would expect given the design of the experiment.

The set of follow-all users likely differs in many ways from the overall population of new Twitter users. 
Importantly, however, the correlational relationship between structural diversity and retention holds in this subset of users, as we discuss below.
Thus, though follow-all users are not representative of new Twitter users, they exhibit the key association between structural diversity and retention that we seek to investigate, and are accordingly a suitable sub-population for our study.

\subsection{Post-treatment selection}

Among those who followed our network recommendations, our treatment conditions induced sizable differences in structural diversity.
But one might worry that the type of user who elected to follow our recommendations may have differed across treatment arms.
Those who followed our recommendations when assigned to the low-diversity condition might not have done so if assigned to the high-diversity condition.
As a result, conditioning on the post-treatment follow-all outcome could in theory bias our estimates of causal effects~\cite{imbens:2015,montgomery:2018}.

There is no complete solution to this problem, 
but we note three factors that mitigate such concerns in our study.
First, approximately the same proportion of eligible users in each condition followed our 20 recommendations:
1.8\% in the control and low-diversity conditions,
2.1\% in the medium-diversity condition, 
and 2.0\% in the high-diversity condition.
Second, as shown in Figure~\ref{fig:balance}, 
the follow-all users in each treatment group are nearly identical on several observable dimensions: 
client platform (e.g., Android or iOS), country, email domain, and language.\footnote{%
This covariate check is only meant
to assess post-treatment bias. 
As we have discussed above, the subset of follow-all users that we consider is not representative of the broader population of new Twitter users.
}
Finally, concerns of selection bias are most salient when outcomes differ across treatment arms, in which case differences may result from selection on unobservables rather than any causal effect of the treatment.
However, as we describe below, we find that outcomes are nearly identical across treatment arms,
and so if a true causal effect exists in our case it would require near-perfect cancellation with the ``follow-all'' selection mechanism to explain our results.

In sum, restricting to the subpopulation of follow-all users likely introduces some selection bias, but the magnitude of the effect appears small and unlikely to affect our qualitative conclusions.
We also note that two common approaches
for dealing with post-treatment selection, Heckman correction~\cite{heckman:1977} and principal stratification~\cite{frangakis2002},
cannot be easily applied in our setting
because
follow-all behavior is not strongly predicted by the available covariates 
and we lack a suitable instrument to circumvent this issue.

\subsection{Design summary}
Though admittedly imperfect, our methodology  helps overcome the key design challenges described above, operationally inducing significant changes to structural diversity while conceptually satisfying a version of exclusion.
Among the follow-all users, we have considerable ability to exogenously vary the structural diversity of their contact networks.
And because follow-all users have a fixed network size of 20, we eliminate network size as a potentially significant factor.
As discussed above, any change to network topology necessarily alters more than structural diversity, but controlling for network size is arguably the most important, and is in line with the approach taken in past observational studies~\cite{ugander2012structural}.
Of course, our restriction to follow-all users limits the generalizability of our results to the full population of Twitter users, but we believe that design choice represents a suitable compromise.

\section{Results}

\subsection{Replicating results of past observational studies}
We begin by examining the relationship between structural diversity and retention on a randomly selected sample of Twitter users who signed up for the service while our experiment was running but who were not themselves part of the experiment. 
As a result, this population contains both eligible and ineligible users, in representative proportions, and all received recommendations from the standard algorithm also used for our control condition.
We throughout consider structural diversity computed on a user's network immediately after joining Twitter, to facilitate direct comparisons to the results from our experiment, in which this initial network structure was randomized.

\begin{figure*}[h!]
\centering
 \includegraphics[height=6.5cm]{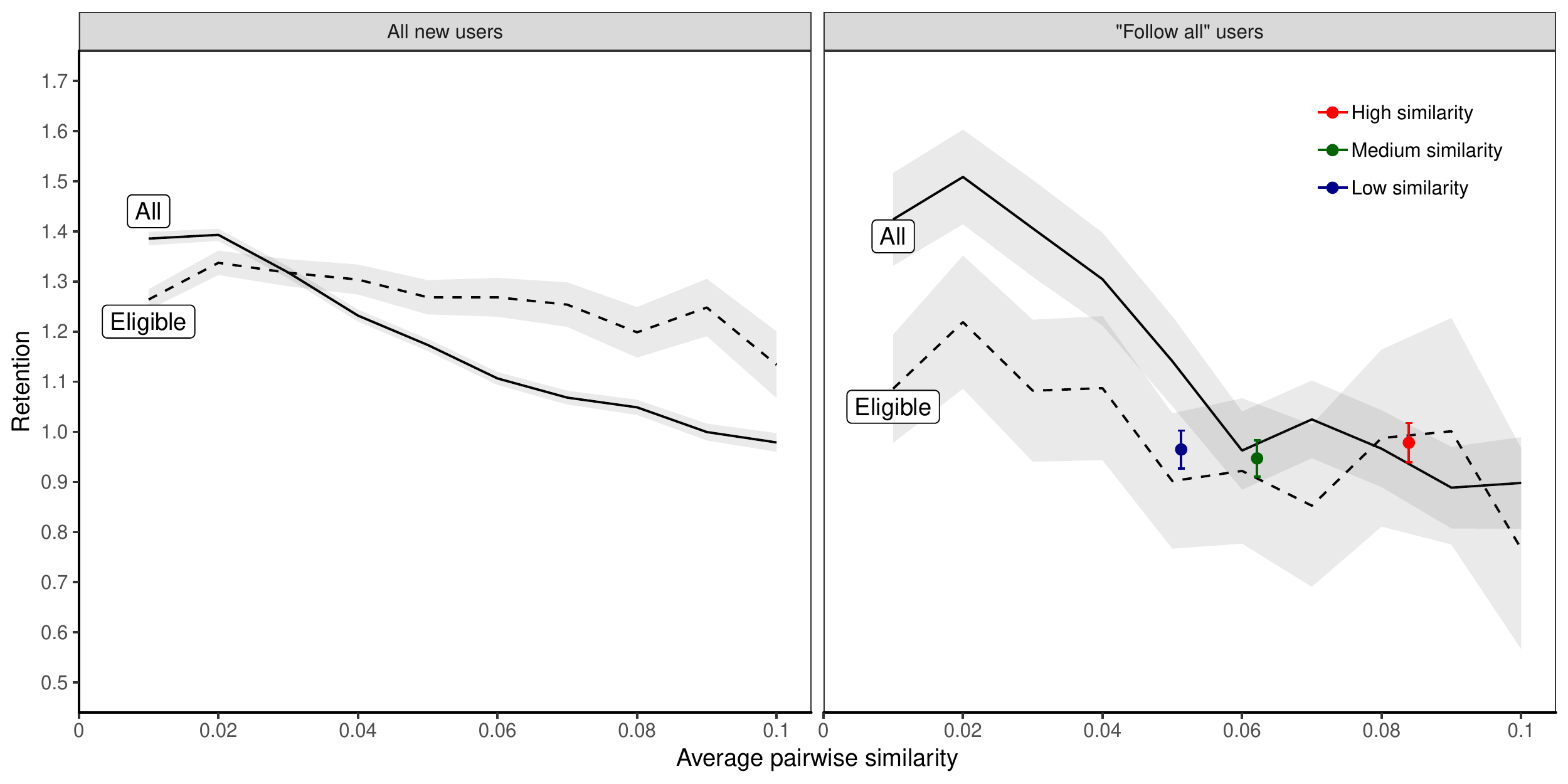}
  \includegraphics[height=6.5cm]{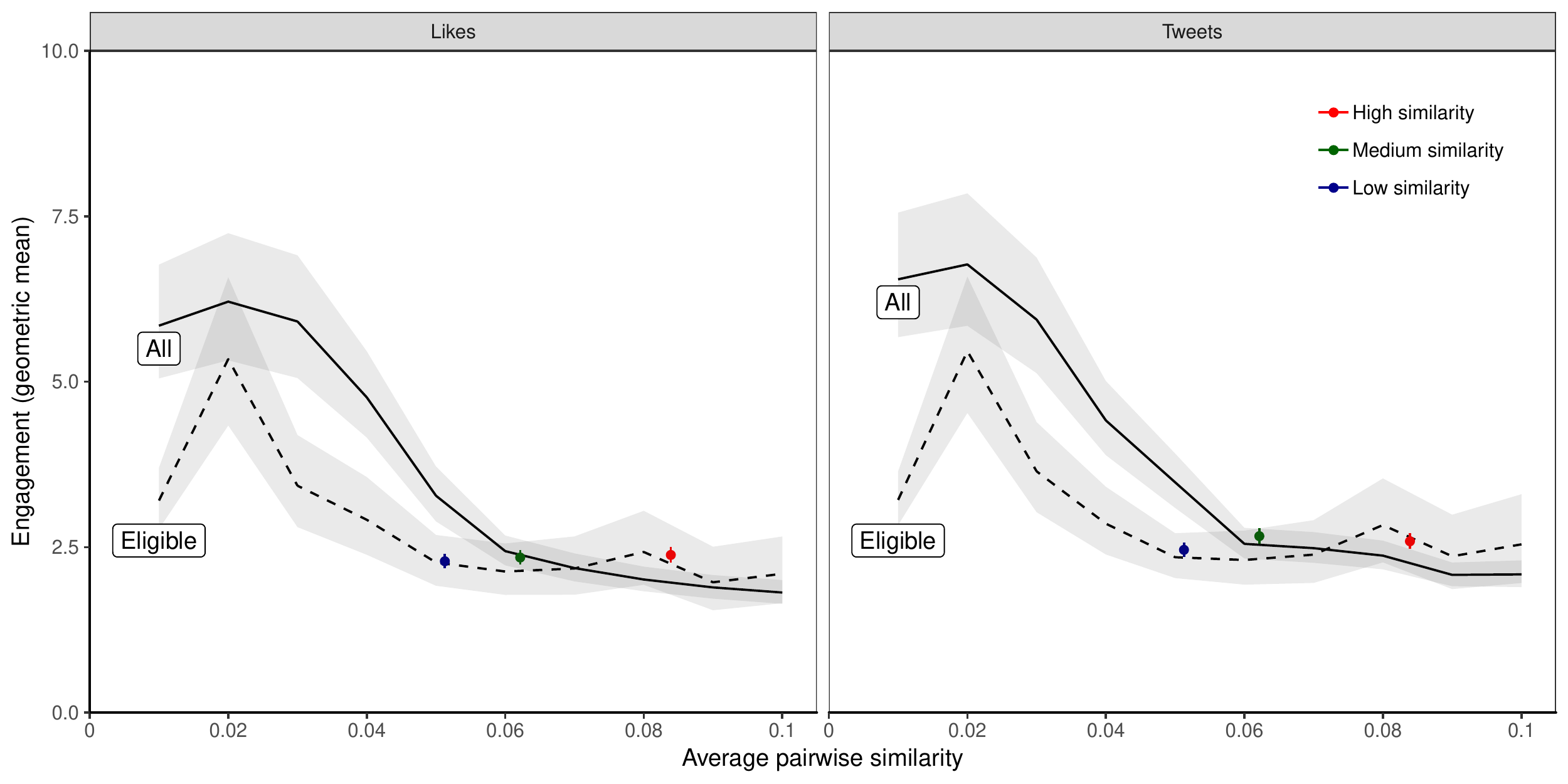}
\caption{Top Left: Among both the full population of Twitter users and the subpopulation of eligible users (from the control condition), structural diversity is correlated with retention, though the effect is attenuated in the latter group. 
Top Right: Analogous results for ``follow all'' users, who have an initial network size of exactly 20.
Despite the observed correlation between structural diversity and retention, retention is comparable across all experiment conditions (colored points, with 95\% confidence intervals), suggesting that structural diversity does not causally affect retention in this case.
Bottom: Alternative engagement metrics (number of likes and number of Tweets by new users) as a function of pairwise similarity, again for the subpopulation of users who follow exactly 20 people. 
We plot the geometric mean of the values after adding 1 to each user's activity measure to account for zeros. 
In all cases, average pairwise similarity is computed based on a user's initial network.
These plots corroborate our primary finding that structural diversity does not appear to be causally connected to engagement.}
\label{fig:results}
\vspace{-2mm}
\end{figure*}

In Figure~\ref{fig:results} (top), users are binned according to the structural diversity of their initial networks, where we use pairwise similarity to quantify diversity.
For each point on the horizontal axis, the vertical axis indicates the fraction of users in that bin who were active on Twitter three months after signing up.
The top-left panel of Figure~\ref{fig:results} is based on all users with an initial network size of at least 10, where this lower limit ensures pairwise similarity can be reliably computed;
the top-right panel is based on the subset of ``follow all'' users, with an initial network size of exactly 20.
In both cases, the solid lines show that retention is positively correlated with structural diversity (i.e., retention decreases with pairwise similarity), corroborating past observational results in other domains.
In accordance with Twitter's reporting policies we show only relative retention rates, with 1.0 in both panels corresponding to the overall retention rate among follow-all users.

Our study differs from past work in the population we study (Twitter users, particularly those who followed our 20 recommendations); 
the outcome we measure (30-day retention);
and our specific measure of structural diversity
(average pairwise similarity). 
Nonetheless, the patterns we see are consistent with the theory that engagement is positively correlated with
structural diversity.
Importantly, this consistency with past work 
indicates that both our setting and our 
design choices are well-suited to 
examining structural diversity.
Had we not been able to replicate past findings, 
one might reasonably worry that idiosyncrasies 
in the particular setting we study were driving the results.

\subsection{Restricting to eligible users}
We next repeat our analysis
for eligible users---the 12\% of the population, as discussed above, 
for whom we could generate recommendations having low, medium, and high structural diversity. 
These eligible users come from our control condition, and they see the standard recommendations Twitter provides.
The results are again shown in Figure~\ref{fig:results} (top panels, dashed lines).
On this eligible subpopulation, we find that the strength of the relationship between retention and structural diversity
is significantly attenuated.
Thus, simply controlling for the \emph{possibility} of selecting into structurally diverse or homogeneous networks, much of the apparent diversity effect on retention disappears. 

The contrast between all users (solid lines) and eligible users (dashed lines) is indeed one of our key findings. 
Once we restrict to a population for whom one can reasonably talk about the causal effects of network structure (i.e., our set of eligible users, who could plausibly enter into low or high diversity contact networks), the relationship between structural diversity and engagement is considerably smaller.
Thus, even without running a randomized trial, 
these preliminary results indicate 
that the observed correlation between structural diversity and retention in the full Twitter population is driven in part by selection rather than causation.

Notably, eligible users have lower overall engagement than the broader population. 
Though we cannot definitively say what is driving that result,
the pattern illustrates the importance of considering only eligible users when testing causal theories of structural diversity. 
The structural diversity of ineligible users, by definition, cannot be easily varied, and so the meaning of causation is ill-posed in that subpopulation. 
Because ineligible users are qualitatively different from those that are eligible---for examle, having higher retention rates---it is important to filter out such users when estimating causal effects.

\subsection{The randomized trial}

Finally, we present the results from our randomized trial.
The three colored points in Figure~\ref{fig:results} (top right) indicate that retention rates for follow-all users are comparable in the low-, medium-, and high-diversity treatment groups.
The locations of these points on the horizontal axis indicate the mean pairwise similarity of each condition, as shown in Figure~\ref{fig:sim-dist}.
The 95\% confidence interval for the difference between the high- and low-similarity treatment groups is $(-0.04, 0.07)$---indicating a lack of statistical significance---where we report these numbers in the units of relative retention rates as in the figure.\footnote{%
Many papers aim to find statistically significant differences between treatment conditions.
In this case, however, the null result is our main finding. Past work has posited that structural diversity and engagement are causally related (based on observational evidence), but we find no evidence of that link in our randomized experiment, as indicated by the lack of statistical significance.
}
We similarly find no statistically significant difference between treatment conditions for the number of Tweets posted or the number of Tweets ``liked'' by new users in their first 30 days on the platform (Figure~\ref{fig:results}, bottom panels).
Thus, at least for the subpopulation we study, we do not find evidence of a causal effect of structural diversity on adoption or engagement.

\section{Conclusion}

Our results corroborate the correlational relationship between structural diversity and behavior found in past observational studies.
But, at least in the setting we consider, we also find 
that this pattern is driven by selection rather than causation.
In particular, we find that the connection between structural diversity and retention (and engagement) is significantly weaker in the subpopulation of eligible users, who might reasonably form either homogeneous or diverse networks.
In the randomized low-, medium- and high-diversity experiment groups we find no significant difference in retention or engagement rates. 
It thus appears that those who organically form diverse networks differ systematically from those who form homogeneous networks in ways that affect behavior.

In establishing these results, we hope to have clarified both
the theoretical underpinnings and the practical challenges of testing for a causal link between network structure and behavior. 
Our work leaves unanswered several important questions, including the extent to which these findings generalize to other domains, populations, and strategies for inducing structural changes.
For example, 
our specific methodological approach required us to focus on a small, non-representative subset of new Twitter users who followed our recommendations, and for whom we could thus exogenously vary structural diversity.
We hope future studies can design experiments to investigate larger subsets of the population.
Further, as implied by our discussion of exclusion, 
there are multiple ways to exogenously vary structural diversity and other methods---including those based on more complete social network information---might yield different results.
Looking forward, we hope our results and our approach spur further exploration of the often subtle causal role that network structure may play in social behavior.
We also hope our results help demonstrate the value of randomized experiments in social network studies~\cite{manski1993identification,shalizi2011}.

\section*{Acknowledgements}
We thank Christopher Chang, Dean Eckles, Avi Feller and Jennifer Pan for their comments. 
We also thank the Recommendations and Notifications team at Twitter, including Bruce Deng, Jerry Jiang, Jinsong Lin, and Zhijun Yin,
for their help with implementing the experiment.
Our study design was reviewed and approved by Stanford University's Institutional Review Board (Protocol \#: 36150).

\balance
\bibliographystyle{aaai}
\bibliography{refs.bib}

\end{document}